\documentclass[prd,aps,twocolumn,amsmath,amssymb,nofootinbib,preprintnumbers]
{revtex4}

\usepackage{graphicx}% Include figure files
\usepackage{dcolumn}% Align table columns on decimal point
\usepackage{bm}% bold math
\usepackage{amsmath}
\usepackage{amsfonts}
\def\ls{\mathrel{\lower4pt\vbox{\lineskip=0pt\baselineskip=0pt
           \hbox{$<$}\hbox{$\sim$}}}}
\def\gs{\mathrel{\lower4pt\vbox{\lineskip=0pt\baselineskip=0pt
           \hbox{$>$}\hbox{$\sim$}}}}
\def\drawbox#1#2{\hrule height#2pt

\hbox{\vrule width#2pt height#1pt \kern#1pt
              \vrule width#2pt}
              \hrule height#2pt}

\def\Asym#1#2{\vcenter{\vbox{\drawbox{#1}{#2}
              \kern-#2pt       % line up boxes
              \drawbox{#1}{#2}}}}

\newcommand{\beq}{\begin{equation}}
\newcommand{\eeq}{\end{equation}}

\begin{document}

%
%\vspace*{2cm}
\title{A Supersymmetric $B-L$ Dark Matter Model \\
and the Observed Anomalies in the Cosmic Rays}

\author{Rouzbeh Allahverdi$^{1}$}
\author{Bhaskar Dutta$^{2}$}
\author{Katherine Richardson-McDaniel$^{1}$}
\author{Yudi Santoso$^{3}$}

\affiliation{$^{1}$~Department of Physics \& Astronomy, University of New Mexico, Albuquerque, NM 87131, USA \\
$^{2}$~Department of Physics, Texas A\&M University, College Station, TX 77843-4242, USA\\
$^{3}$~Institute for Particle Physics Phenomenology, Department of Physics, University of Durham, Durham DH1 3LE, UK}

%\date{Dec 11, 2008}

\begin{abstract}

We propose a simple model of supersymmetric dark matter that can explain recent results from PAMELA and ATIC experiments. It is based on a $U(1)_{B-L}$ extension of the minimal supersymmetric standard model. The dark matter particle is a linear combination of the $U(1)_{B-L}$ gaugino and Higgsino partners of Higgs fields that break the $B-L$ around one TeV. The dominant mode of dark matter annihilation is to the lightest of the new Higgs fields, which has a mass in the GeV range, and its subsequent decay mainly produces taus or muons by the virtue of $B-L$ charges. This light Higgs also results in Sommerfeld enhancement of the dark matter annihilation cross section, which can be $\gs 10^3$. For a dark matter mass in the $1-2$~TeV range, the model provides a good fit to the PAMELA data and a reasonable fit to the ATIC data. We also briefly discuss the prospects of this model for direct detection experiments and the LHC.

\end{abstract}

MIFTP-08-31, IPPP/08/90, DCPT/08/180 \\ December, 2008
\maketitle
%%%%%%%%%%%%%%%%%%%%%%%%%%%%%%%%%%%%%%%%%%%%%%%%%%%%%%%%%%%%%%%%%%%%%%%%%%%%%%%
%\section{Introduction}

One of the major problems at the interface of particle physics and cosmology is dark matter and its identity. There are various lines of evidence supporting the existence of dark matter in the universe, and it is well established that particle physics can explain dark matter in the form of weakly interacting massive particles~\cite{Silk}.
The standard scenario is thermal dark matter in which the dark matter relic abundance, as precisely measured by cosmic microwave background (CMB) experiments~\cite{WMAP5}, is determined from the freeze out of the dark matter annihilation in the early universe.
Supersymmetry is a front runner candidate for physics beyond the standard model (SM). It addresses the hierarchy problem of the SM and has a natural dark matter candidate in the form of the lightest supersymmetric particle (LSP).
It has been known that in supersymmetric models, a neutralino LSP can have the thermal relic abundance required for dark matter~\cite{EHNOS}.

There are major experimental efforts for direct and indirect detection of dark matter particle beside the gravitational effect that it has on the universe. Indirect detection investigates final states (photons, anti-particles, neutrinos) from the annihilation of dark matter through astrophysical observations, while the direct detection probes the scattering of the dark matter particle off nuclei in the dark matter detectors. The PAMELA satellite is an indirect experiment that has recently published results on cosmic ray flux measurements. The data show an excess of positrons at energies above 10~GeV~\cite{PAMELA}, while no excess of anti-proton flux is observed~\cite{PAMELA-antiproton}. The publication shows results up to 100~GeV and the experiment is expected to get data up to $\sim 270$~ GeV. There is also new data from the ATIC (a balloon experiment) where one observes excess in $e^{+} + e^{-}$ spectrum with a peak around 600~GeV~\cite{ATIC}. The backgrounds are nominal for both of these effects.
Another balloon experiment, the PPB-BETS~\cite{pep}, also reports excess in the $e^+ + e^-$ energy spectrum between 500-800 GeV. However, the excess is based on a few data-points that are not quite consistent with ATIC data.
While there could be astrophysical explanations for these anomalies~\cite{pulsars}, it is natural to ask whether they can be attributed to the effect of dark matter annihilation in the halo.

Model-independent analysis shows that the annihilation cross section required to explain the positron excess exceeds the canonical value required by relic density $\sim 3 \times 10^{-26}$~cm$^3$/s by at least an order of magnitude~\cite{Vernon}. In the usual minimal Supergravity (mSUGRA) model, the situation is further complicated since dark matter annihilation to fermions in that model is $P$-wave suppressed, implying a much smaller annihilation cross section today as compared to that at the freeze out time. Even in the best case scenarios an astrophysical boost factor of $10^3-10^4$ is then needed to explain the positron excess~\cite{Lars}, which might be difficult to obtain based on the recent analysis on substructures(see e.g.~\cite{halostructure}). Moreover, in order to explain both the positron and anti-proton data, dark matter annihilation must dominantly produce leptons as direct products~\cite{Strumia,Salati}. There have been proposals~\cite{Strumia,Nima} (also
see~\cite{Other,decaying}) for new dark matter models in which the dark matter candidate belongs to a hidden sector. Acceptable thermal relic density is obtained via new gauge interactions, dark matter annihilation today is enhanced via Sommerfeld effect~\cite{Sommerfeld} due to existence of light bosons, and the annihilation mainly produces lepton final states. This set up can explain PAMELA data for dark matter mass of a few hundred GeV without needing a large boost factor, and ATIC data for larger masses~\cite{Weiner}.

In this paper we provide a concrete model to explain the recently measured anomalies in the cosmic ray. Our model is a simple extension of the minimal supersymmetric standard model (MSSM) that includes a gauged $U(1)_{B-L}$~\cite{mohapatra}. The $B-L$ extension is very well motivated since it automatically implies the existence of three right-handed (RH) neutrinos through which one can explain the neutrino masses and mixings. It has also been shown that this model can explain inflation~\cite{Inflation}. The model contains a new gauge boson $Z^{\prime}$, two new Higgs fields $H^{\prime}_1$ and $H^{\prime}_2$, and their supersymmetric partners.

%
%%%%%%%%%%%%%%%%%%%%
%\section{Model}
%
The $B-L$ charge assignments are shown in Table 1. The superpotential is
\begin{equation}
W = W_{\rm MSSM} + W_N + \mu^\prime {H}^{\prime}_1 {H}^{\prime}_2
\end{equation}
where $W_N$ is the superpotential containing RH neutrinos, and $\mu^\prime$ is the new Higgs mixing parameter.
The new Higgs fields do not have renormalizable coupling to lepton and quark fermions as a result of their $B-L$ charges. The $U(1)_{B-L}$ symmetry is broken by the VEV of the new Higgs bosons, $v_1^\prime \equiv \langle H^{\prime}_1 \rangle$ and $v_2^\prime \equiv \langle H^{\prime}_2 \rangle$. This gives a mass $m_{Z^{\prime}}$ to $Z^{\prime}$, where $m^2_{Z^{\prime}} = (27/4) g^2_{B-L} (v_1^{\prime 2} + v_2^{\prime 2})$, with $g_{B-L}$ being the $B-L$ gauge coupling. We have three physical Higgs states, the lightest of which $\phi$ has a mass $m_{\phi}$ which is related to $m_{Z^{\prime}}$ through $m^2_{\phi} < m^2_{Z^{\prime}} \cos^2 2 \beta^{\prime}$, where $\tan \beta^{\prime} \equiv v_2^{\prime} / v_1^{\prime}$. For $\tan \beta^{\prime} \approx 1$ we have $m_{\phi} \ll m_{Z^{\prime}}$. The other two Higgs states, $\Phi$ and ${\cal A}$, are heavy and have masses comparable to $m_{Z^{\prime}}$. Note that unlike the MSSM case, radiative corrections to the Higgs quartic coupling do not lift $m_{\phi}$ because $H^{\prime}_1$ and $H^{\prime}_2$ are not coupled to fermions. Note that although the particle content in our model is similar to the one in~\cite{khalil}, our set up is different.
\begin{table}[tbp]
\center
\begin{tabular}{|c||c||c|c|c|c|c|c|c|c|}\hline
{\rm Fields} & $Q$ & $Q^c$ & $L$ & $L^c$ & $N$ & $N^c$ & $H^{\prime}_1$ & $H^{\prime}_2$ \\ \hline
$Q_{B-L}$ & 1/6 & -1/6 & -1/2 & 1/2 & -1/2 & 1/2 & 3/2 & -3/2 \\ \hline
\end{tabular}
\caption{The $B-L$ charges of the fields. Here $Q$, $L$ and $N$ represent quarks, leptons, and RH neutrinos respectively; while $H^{\prime}_1$ and $H^{\prime}_2$ are the two new Higgs fields. The MSSM Higgs fields have zero $B-L$ charges and are not shown in the table.}
%\label{}
\end{table}
Assuming that supersymmetric particles in the MSSM sector are heavier than those in the $U(1)_{B-L}$ sector, the dark matter in this model arises from the new sector. The dark matter particle, denoted by $\chi^0_1$, is the lightest of the three new neutralinos $\chi^0_i$ (not to be confused with the MSSM neutralinos which we do not discuss in this paper). It is a linear combination of the $U(1)_{B-L}$ gaugino ${\widetilde Z}^{\prime}$ and the two Higgsinos ${\widetilde H}^{\prime}_1$, ${\widetilde H}^{\prime}_2$.
The dark matter thermal relic abundance is dictated by the annihilation of the lightest neutralino $\chi_1^0$ into a pair of $\phi$ via $s$-channel exchange of $\phi$ and  $\Phi$, and $t$-channel exchange of $\chi^0_i$. (There is also subdominant annihilation processes to $f \bar{f}$ via $s$-channel $Z^\prime$ exchange, and $t$-channel sfermion exchange.) The annihilation to $\phi \phi$ is not $P$-wave suppressed since the final state particles are bosons. Hence the (perturbative) annihilation cross-section does not change between the time of freeze-out and the present time.

The $\phi$ subsequently decays into fermion-antifermion pairs via a one-loop diagram containing two $Z^{\prime}$ bosons. The decay rate is given by:
\begin{equation} \label{phidec}
\Gamma(\phi \to f {\bar f}) = \frac{C_f}{2^6 \pi^5} \frac{g_{B-L}^6 Q^4_{f} Q^2_{\phi} m^5_{\phi} m^2_f}{m^6_{Z^{\prime}}} \left(\frac{1}{2} - \frac{2 m^2_f}{m^2_{\phi}} \right),
\end{equation}
where $Q_{\phi}$ and $Q_f$ are the $B-L$ charges of $\phi$ and the final state fermion respectively, $m_f$ is the fermion mass, and $C_f$ denotes color factor. (The Higgs can also decay into four fermion final states via two virtual $Z^{\prime}$, but this decay mode is suppressed by two orders of magnitude compared to the two fermion final states.) Decays to neutrinos are much suppressed because of the negligible neutrino masses. Since $B-L$ charge of leptons is three times larger than that of quarks, the leptonic branching ratio is about $27$ times larger than that for quarks of comparable mass. We note that $m_{\phi}$ can be controlled by the VEVs of the new Higgs fields and for comparable VEVs, i.e. for $\tan \beta^{\prime} \approx 1$, it can be very small without any tuning on the soft masses in the Higgs sector. We can choose this mass to be between $O(1)$~GeV and 10~GeV. For $2 m_\tau < m_{\phi} < 2 m_b$ the dominant decay mode is to $\tau^{-} \tau^{+}$ final state. If $m_\phi$ is slightly less than $2 m_\tau$, $\phi$ can decay either to $c {\bar c}$ or $\mu^{-} \mu^{+}$ with comparable branching ratios. It is possible to reduce the $\phi$ mass further to be below $2 m_c$, and make $\mu^{-} \mu^{+}$ final state the dominant decay mode.

The annihilation cross section at the present time has Sommerfeld enhancement as a result of the attractive force between dark matter particles due to the light Higgs boson exchange. The Higgs coupling to dark matter, $h \phi {\bar {\chi}^0_1} \chi^0_1$, leads to an attractive potential $V(r) = -\alpha (e^{-m_{\phi}r}/r)$ in the non-relativistic limit, where $\alpha \equiv h^2/\pi$. Note that $\alpha$ is larger than the usual definition of fine structure constant because of the Majorana nature of $\chi^0_1$. Since the neutralinos are traveling with non-relativistic speed $v \sim 10^{-3} c$ today, the Sommerfeld effect is much more important now than at the time of freeze out. The Sommerfeld enhancement in our model can be $\gs 10^{3}$. We can therefore explain the PAMELA data without requiring any boost factor.

%%%%%%%%%%%%%%%%%%%%%%%%%%%%%%%%%%
%\section{Results}

To show that our model can explain the PAMELA data, we pick random parameters and generate point models. We then calculate the relic density and the Sommerfeld enhancement factor for each of these models. We use reasonable values for the parameters, i.e. $\tan \beta^\prime \approx 1$, $m_{Z^\prime} > 1.5$~TeV, $\mu^\prime = 0.5 - 1.5$~TeV, soft masses for the Higgs fields $m_{H_{1,2}^\prime} = 200-600$~GeV, and soft gaugino mass $M_{\widetilde{Z}^\prime} = 200-600$~GeV. We use $g_{B-L} \sim 0.45$, which is in concordance with unification of the gauge couplings (we need to use a normalization factor $\sqrt{3/2}$ for unification). The $Z^{\prime}$ mass used in the calculation obeys the LEP and the Tevatron bounds~\cite{tev,carena} for our charge assignments.

\begin{figure}
%\vspace*{0.5cm}
\begin{center}
\includegraphics[width=8.0cm]{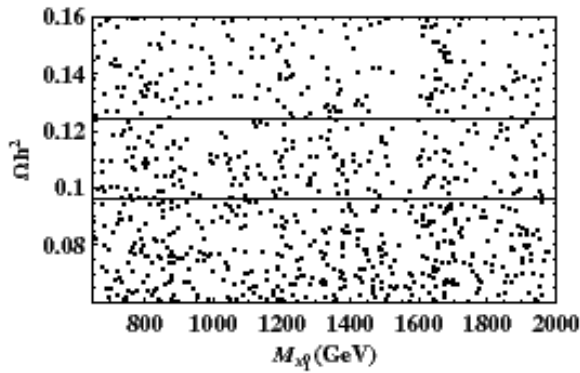}
\end{center}
\vskip -0.25in
\caption{We show the relic density and the neutralino mass for model points generated by varying the parameters mentioned in the text. \label{ohsqvsm}}
\label{model}
\end{figure}

\begin{figure}
%\vspace*{0.5cm}
\begin{center}
\includegraphics[width=8.0cm]{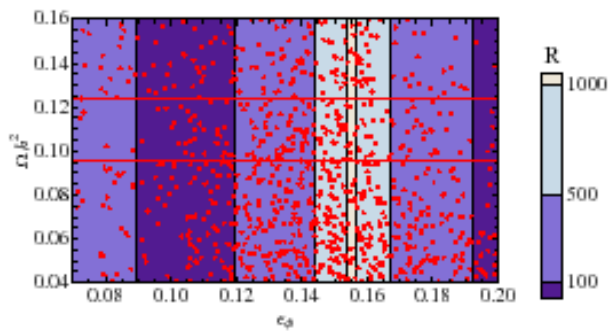}
\end{center}
\vskip -0.25in
\caption{We show relic density as a function of $\epsilon_\phi$. We show different ranges for the Sommerfeld enhancement factor $R$ by shades.\label{ohsqvsephi}}
\end{figure}

In Figure~\ref{ohsqvsm}, we show the relic density and the dark matter mass for different points. The horizontal band shows the acceptable range for relic density according to the latest CMB data~\cite{WMAP5}. In Figure~\ref{ohsqvsephi} we show the possible enhancement that can be obtained for these points in term of $\epsilon_\phi \equiv m_{\phi}/\alpha M_{\chi^0_1}$. Note that for points that satisfy the relic density constraint thereare
many that have enhancement factor $\geq 10^3$,
corresponding to $\epsilon_\phi = 0.153$ to $0.157$, within the whole
range of $M_{\chi^0_1}$ shown in Figure~\ref{ohsqvsm}. In addition, we note that the lifetime of $\phi$ for these points is found to be $\tau_{\phi} \sim 10^{-4}-10^{-3}$~seconds from Eq. (2). This is short enough to escape the tightest bounds from big bang nucleosynthesis (BBN)~\cite{BBN}.

We select models that satisfy the dark matter relic density. We then use {\tt DarkSUSY}-5.0.2~\cite{darksusy} to calculate the positron flux from dark matter annihilation. Note that each pair annihilation in our model produces 2 $\phi$'s that yield four fermions upon their decay. For this reason, we generally need a heavier neutralino compared to models in which the pair annihilation directly produces two fermions. We normalize the positron fraction by a factor $k_b = 1.11$~\cite{BE}. There are theoretical uncertainties in the positron cosmic ray flux calculation due to the assumptions about the dark matter halo profile and the cosmic ray propagation model. Here we assume a NFW profile~\cite{NFW} and MED parameters for the propagation as defined in~\cite{Delahaye}.

\begin{figure}
%\vspace*{1.0cm}
\begin{center}
\vskip -0.9in
\includegraphics[width=.48\textwidth]{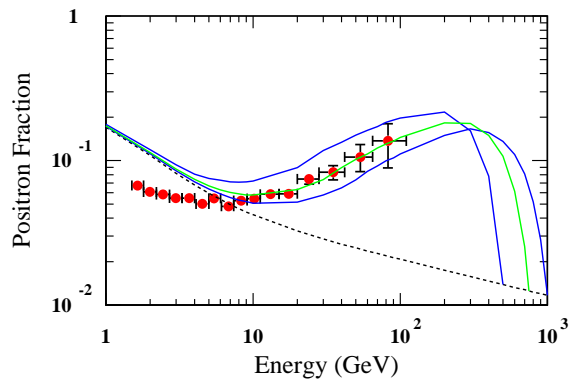}
\end{center}
\vskip -0.65in
\caption{We show a fit to the PAMELA data when the $\phi$ decays mostly to taus for neutralino masses to be 1, 1.5, and 2~TeV (from top to bottom), with enhancement factor $10^3$.}
\label{pamelatau}
\end{figure}

In Figure~\ref{pamelatau}, we show our fit to the PAMELA data for $M_{\chi^0_1} = 1, 1.5$ and 2~TeV, for $\tau^- \tau^+$ final state case. We see that the fit is very good for a neutralino mass around 1.5 TeV. We have chosen a point where the enhancement factor is $10^3$. In general, for larger enhancement factors we can fit the data with a larger neutralino mass. For an enhancement factor of $10^4$, we can obtain a good fit for a dark matter mass around 5~TeV. On the other hand, smaller masses require a smaller enhancement factor. However, for this tau case the fit with a smaller mass is not good because the spectrum is too soft. The antiproton data is still satisfied with this large enhancement factor since the leptonic branching ratio is much larger ($\sim 27$) compared to the quark branching ratios.

\begin{figure}
%\vspace*{1.0cm}
\begin{center}
\vskip -0.8in
\includegraphics[width=.48\textwidth]{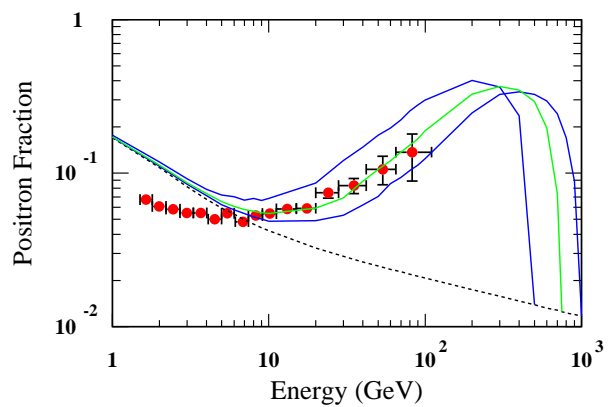}
\end{center}
\vskip -0.65in
\caption{Same as Fig.~\ref{pamelatau} but for the case of final state muons.}
\label{pamelamu}
\end{figure}

In Figure~\ref{pamelamu} we show the plot with $\mu^{-} \mu^{+}$ final states. In order for the Higgs to mainly decay to muons, we need to make $m_{\phi}$ smaller than twice the charm mass. However, the lifetime of the Higgs becomes larger than a second in this case, which will be problematic for BBN~\cite{BBN}. The Higgs lifetime can be reduced if we increase the $B-L$ charges by a factor of two. On the other hand, the positron spectrum in the muon case is harder than that in the tau case, and we can also fit the PAMELA data with a 500~GeV neutralino mass, and an enhancement factor of around 100.

\begin{figure}
%\vspace*{1.0cm}
\begin{center}
\vskip -0.8in
\includegraphics[width=.48\textwidth]{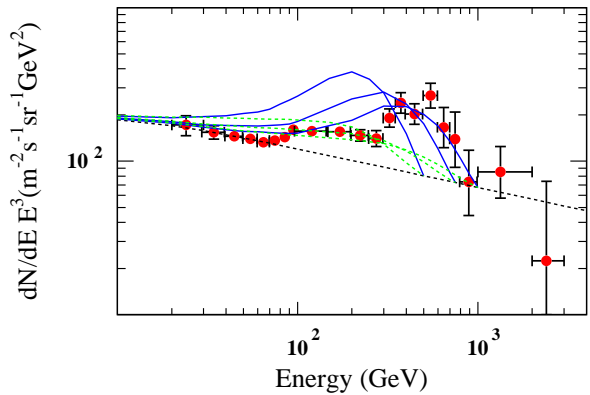}
\end{center}
\vskip -0.65in
\caption{We show a fit to the ATIC data when $\phi$ decays to muons mostly (solid line) and taus mostly(dashed line) for the lightest neutralino masses to be 1, 1.5 and 2~TeV}
\label{aticmutau}
\end{figure}

In Figure~\ref{aticmutau} we show the fit to the ATIC data by using muon and tau final states. We normalize the background to fit with the data at smaller energies. We again use enhancement factor of $10^3$. We see that the muon final states give a better fit than the tau final states.
However, we can see that the simultaneous fit to both ATIC and PAMELA is not satisfactory. In particular, it is more difficult to fit the ATIC data with our model. If we use a large enhancement factor, the spectrum at smaller energies will also be lifted up. From the ATIC data plot~\cite{ATIC}, there seems to be some missing signals at some energies which results in the jagged curve. This can also indicate more than one source for the excess. In any case, we need to wait for verification from future experiments. It is interesting to point out that there is a proposal for an upgraded version of ATIC, called ECAL~\cite{ECAL}, which should have much improved background rejection power and higher resolution. We will be able identify the model parameter space by combining results from all these experiments.

%%%%%%%%%%%%%%%%%%%%%%%%%%%%%%%%%%
%\section{Other Aspects}

We would also like to comment on other phenomenological aspects of the model.
%
%\subsection{Direct detection}
%
The leading order interaction of dark matter particle $\chi^0_1$ with quarks is via squark exchange in the $t$-channel. The new Higgs fields couple to the quarks at one-loop level, and hence interactions via Higgs exchange are suppressed (even after Sommerfled enhancement is taken into account). As a result, the cross section for spin-independent interactions is also very small, well below $10^{-10}$ pb, and hence beyond the reach of direct detection experiments.
We note that since left and right quarks have the same $B-L$ charge, and $\chi^0_1$ is a Majorana particle, there will be strictly no spin-dependent interactions between dark matter and ordinary matter in this model.
The model however has a great potential to be observed with the Fermi Satellite experiment. The Sommerfeld enhancement would still be responsible for giving rise to a higher rate of photons in the cosmic gamma ray background.

%%%%%%%%%%%%%%%%%%%%%%%%%%%%%%%%%%%%%%%%%%%%%%%%%%%%%%%%%
%\subsection{LHC}

At the LHC, the $Z^{\prime}$ can be produced. However, the new light Higgs $\phi$ will decay outside of the detector, hence will be missed, because of its long life time ($\sim 10^{-4}$ sec).
Thus we have another source of missing energy signal in this model. We also note that there are 7 neutralinos in this model, compared to four in the MSSM, while the number of charginos is still two. Therefore, using the end point analysis~\cite{endpoint}, one can find many neutral states.

%%%%%%%%%%%%%%%%%%%%%%%%%%%%%%%%%%%%%%%%%%%%%%%%%%%%%%%%
%\section{Conclusion}

{\bf Acknowledgement-} The authors would like to thank S.~Bornhauser, M.~Cvetic, P.~Gondolo, R.~Mahapatra, D.~Toback, L.~Wang and J.P.~Wefel for useful discussions and communications. The work of BD is supported in part by DOE  grant DE-FG02-95ER40917.

%%%%%%%%%%%%%%%%%%%%%%%%%%%%%%%%%%%%%%%%%%%%%%%%%%%%%%%%%


\begin{thebibliography}{99}

\bibitem{Silk}
G.~Bertone, D.~Hooper and J.~Silk, Phys.~Rept. {\bf 405}, 279 (2005).

\bibitem{WMAP5}
E.~Komatsu, {\it et al.}, arXiv:0803.0547.
% [astro-ph].

\bibitem{EHNOS}
J.R.~Ellis, J.S.~Hagelin, D.V.~Nanopoulos, K.A.~Olive and M.~Srednicki, Nucl.\ Phys.\  B {\bf 238}, 453 (1984).


\bibitem{PAMELA}
O.~Adriani, {\it et al.}, arXiv:0810.4995.
% [astro-ph].

\bibitem{PAMELA-antiproton}
O.~Adriani, {\it et al.}, arXiv:0810.4994.
% [astro-ph].

\bibitem{ATIC}
J.~Chang, {\it et al.}, Nature 456, 362 (2008).



\bibitem{pep}
S.~Torii, {\it et al.}, arXiv:0809.0760.
% [astro-ph].


\bibitem{pulsars}
D.~Hooper, P.~Blasi and P.D.~Serpico, arXiv:0810.1527.
% [astro-ph].



\bibitem{Vernon}
V.~Barger, W.-Y.~Keung, D.~Marfatia and G.~Shaughnessy, arXiv:0809.0162.
% [hep-ph].

\bibitem{Lars}
L.~Bergstrom, T.~Bringmann and J.~Edsjo, arXiv:0808.3725.
% [astro-ph].

\bibitem{halostructure}
N.~Afshordi, R.~Mohayaee and E.~Bertschinger, arXiv:0811.1582.
% [astro-ph].

\bibitem{Strumia}
M.~Cirelli, M.~Kadastik, M.~Raidal and A.~Strumia; arXiv:0809.2409.
% [hep-ph].

\bibitem{Salati}
F.~Donato, D.~Maurin, P.~Brun, T.~Delahaye and P.~Salati, arXiv:0810.5292.
% [astro-ph].

\bibitem{Nima}
N.~Arkani-Hamed, D.P.~Finkbeiner, T.~Slatyer and N.~Weiner, arXiv:0810.0713.
% [hep-ph].


\bibitem{Other}
%\bibitem{Grajek:2008jb}
P.~Grajek, G.~Kane, D.J.~Phalen, A.~Pierce and S.~Watson,
%``Neutralino Dark Matter from Indirect Detection Revisited,''
arXiv:0807.1508;
%[hep-ph];
%%CITATION = ARXIV:0807.1508;%%
%\bibitem{Pospelov:2008jd}
M.~Pospelov and A.~Ritz,
%``Astrophysical Signatures of Secluded Dark Matter,''
arXiv:0810.1502;
% [hep-ph];
%%CITATION = ARXIV:0810.1502;%%
%\bibitem{Cholis:2008qq}
I.~Cholis, D.P.~Finkbeiner, L.~Goodenough and N.~Weiner,
%``The PAMELA Positron Excess from Annihilations into a Light Boson,''
arXiv:0810.5344;
% [astro-ph];
%%CITATION = ARXIV:0810.5344;%%
%\bibitem{Huh:2008vj}
J.H.~Huh, J.E.~Kim and B.~Kyae,
%``Two dark matter components in N_{DM}MSSM and PAMELA data,''
arXiv:0809.2601;
% [hep-ph];
%%CITATION = ARXIV:0809.2601;%%
%\bibitem{Fairbairn:2008fb}
M.~Fairbairn and J.~Zupan,
%``Two component dark matter,''
arXiv:0810.4147;
% [hep-ph];
%%CITATION = ARXIV:0810.4147;%%
%\bibitem{Nelson:2008hj}
A.E.~Nelson and C.~Spitzer,
%``Slightly Non-Minimal Dark Matter in PAMELA and ATIC,''
arXiv:0810.5167;
% [hep-ph];
%%CITATION = ARXIV:0810.5167;%%
%\bibitem{Feldman:2008xs}
D.~Feldman, Z.~Liu and P.~Nath,
%``PAMELA Positron Excess as a Signal from the Hidden Sector,''
arXiv:0810.5762;
% [hep-ph];
%%CITATION = ARXIV:0810.5762;%%
%\bibitem{Nomura:2008ru}
Y.~Nomura and J.~Thaler,
%``Dark Matter through the Axion Portal,''
arXiv:0810.5397;
% [hep-ph];
%%CITATION = ARXIV:0810.5397;%%
%\bibitem{Ishiwata:2008cv}
K.~Ishiwata, S.~Matsumoto and T.~Moroi,
%``Cosmic-Ray Positron from Superparticle Dark Matter and the PAMELA
%Anomaly,''
arXiv:0811.0250;
% [hep-ph];
%%CITATION = ARXIV:0811.0250;%%
%\bibitem{Fox:2008kb}
P.J.~Fox and E.~Poppitz,
%``Leptophilic Dark Matter,''
arXiv:0811.0399;
% [hep-ph];
%%CITATION = ARXIV:0811.0399;%%
%\bibitem{Harnik:2008uu}
R.~Harnik and G.D.~Kribs,
%``An Effective Theory of Dirac Dark Matter,''
arXiv:0810.5557;
% [hep-ph];
%%CITATION = ARXIV:0810.5557;%%
%\bibitem{Chen:2008md}
C.R.~Chen, F.~Takahashi and T.T.~Yanagida,
%``High-energy Cosmic-Ray Positrons from Hidden-Gauge-Boson Dark Matter,''
arXiv:0811.0477;
  % [hep-ph].
  %%CITATION = ARXIV:0811.0477;%%
%\bibitem{Baek:2008nz}
S.~Baek and P.~Ko,
%``Phenomenology of $U(1)_{L_\mu - L_\tau}$ charged dark matter at PAMELA and
%colliders,''
arXiv:0811.1646.
  % [hep-ph].
  %%CITATION = ARXIV:0811.1646;%%




\bibitem{decaying}
%\bibitem{Chen:2008qs}
C.R.~Chen, M.M.~Nojiri, F.~Takahashi and T.T.~Yanagida,
%``Decaying Hidden Gauge Boson and the PAMELA and ATIC/PPB-BETS Anomalies,''
arXiv:0811.3357;
% [astro-ph];
%%CITATION = ARXIV:0811.3357;%%
%\bibitem{Hamaguchi:2008rv}
K.~Hamaguchi, E.~Nakamura, S.~Shirai and T.T.~Yanagida,
%``Decaying Dark Matter Baryons in a Composite Messenger Model,''
arXiv:0811.0737;
% [hep-ph];
%%CITATION = ARXIV:0811.0737;%%
%\bibitem{Yin:2008bs}
P.f.~Yin, Q.~Yuan, J.~Liu, J.~Zhang, X.j.~Bi and S.h.~Zhu,
%``PAMELA data and leptonically decaying dark matter,''
arXiv:0811.0176;
% [hep-ph];
%%CITATION = ARXIV:0811.0176;%%
%\bibitem{Ibarra:2008jk}
A.~Ibarra and D.~Tran,
%``Decaying Dark Matter and the PAMELA Anomaly,''
arXiv:0811.1555.
  % [hep-ph].
  %%CITATION = ARXIV:0811.1555;%%





\bibitem{Sommerfeld}
A.~Sommerfeld, Annalen der Physik, {\bf 403}, 257 (1931).

\bibitem{Weiner}
I.~Cholis, G.~Dobler, D.P.~Finkbeiner, L.~Goodenough and N.~Weiner, arXiv:0811.3641.
% [astro-ph].

\bibitem{mohapatra}
%\bibitem{Mohapatra:1980qe}
  R.~N.~Mohapatra and R.~E.~Marshak,
  %``Local B-L Symmetry Of Electroweak Interactions, Majorana Neutrinos And
  %Neutron Oscillations,''
  Phys.\ Rev.\ Lett.\  {\bf 44}, 1316 (1980)
  [Erratum-ibid.\  {\bf 44}, 1643 (1980)].
  %%CITATION = PRLTA,44,1316;%%

\bibitem{Inflation}
R.~Allahverdi, A.~Kusenko and A.~Mazumdar, JCAP {\bf 0707}, 018 (2007); R.~Allahverdi, B.~Dutta and A.~Mazumdar, Phys. Rev. Lett. {\bf 99}, 261301 (2007).

\bibitem{khalil}
  S.~Khalil and H.~Okada,
  %``Dark Matter in B-L Extended MSSM Models,''
  arXiv:0810.4573 [hep-ph].
  %%CITATION = ARXIV:0810.4573;%%

\bibitem{tev}
T.~Aaltonen {\it et al.}  [CDF Collaboration],
  %``Search for new physics in high mass electron-positron events in $p \bar{p}$
  %collisions at $\sqrt{s}$ = 1.96-TeV,''
  Phys.\ Rev.\ Lett.\  {\bf 99}, 171802 (2007).
  [arXiv:0707.2524 [hep-ex]].
  %%CITATION = PRLTA,99,171802;%%
  
\bibitem{carena}
M.~S.~Carena, A.~Daleo, B.~A.~Dobrescu and T.~M.~P.~Tait,
  %``Z' gauge bosons at the Tevatron,''
  Phys.\ Rev.\  D {\bf 70}, 093009 (2004)
  [arXiv:hep-ph/0408098].
  %%CITATION = PHRVA,D70,093009;%%
  

\bibitem{BBN}
M.~Kawasaki, K.~Kohri and T.~Moroi, Phys. Rev. D {\bf 71}, 083502 (2005).

\bibitem{darksusy}
%\bibitem{Gondolo:2004sc}
P.~Gondolo, J.~Edsjo, P.~Ullio, L.~Bergstrom, M.~Schelke and E.A.~Baltz,
%``DarkSUSY: Computing supersymmetric dark matter properties numerically,''
JCAP {\bf 0407}, 008 (2004).
  %[arXiv:astro-ph/0406204].
  %%CITATION = JCAPA,0407,008;%%%

\bibitem{BE}
%\bibitem{Baltz:1998xv}
E.A.~Baltz and J.~Edsjo,
%``Positron Propagation and Fluxes from Neutralino Annihilation in the Halo,''
Phys.\ Rev.\  D {\bf 59}, 023511 (1999).
  %[arXiv:astro-ph/9808243].
  %%CITATION = PHRVA,D59,023511;%%


\bibitem{NFW}
%\bibitem{Navarro:1995iw}
J.F.~Navarro, C.S.~Frenk and S.D.M.~White,
%``The Structure of Cold Dark Matter Halos,''
Astrophys.\ J.\  {\bf 462}, 563 (1996).
  %[arXiv:astro-ph/9508025].
  %%CITATION = ASJOA,462,563;%%

\bibitem{Delahaye}
%\bibitem{Delahaye:2007fr}
T.~Delahaye, R.~Lineros, F.~Donato, N.~Fornengo and P.~Salati,
%``Positrons from dark matter annihilation in the galactic halo: theoretical
%uncertainties,''
Phys.\ Rev.\ D {\bf 77}, 063527 (2008).
  %[arXiv:0712.2312 [astro-ph]].
  %%CITATION = PHRVA,D77,063527;%%


\bibitem{ECAL}
T.G.~Guzik, {\it et al.},
% The Electron Calorimeter (ECAL) Long Duration Ballon Experiment,
in ICRC 2007 Proceedings.

\bibitem{endpoint}
I.~Hinchliffe, F.E.~Paige, M.D.~Shapiro, J.~Soderqvist and W.~Yao,
  %``Precision SUSY measurements at LHC,''
Phys.\ Rev.\ D {\bf 55}, 5520 (1997).
  %[arXiv:hep-ph/9610544].
  %%CITATION = PHRVA,D55,5520;%%



\end{thebibliography}
\end{document}